\documentclass{./emulateapj}

\usepackage{color,epsfig,graphics,graphicx,./pdfpages,./psfig}
\usepackage{natbib}
\definecolor{brown}{rgb}{0.6,0.4,0.2} 
\definecolor{purple}{rgb}{0.5,0,0.5} 

\newcommand{\oif}{[O\,I]}

\newcommand{\oiiif}{\ion{[O}{3}]}  
\newcommand{\oivf}{\ion{[O}{4}]}

\newcommand{\kms}{km~s$^{-1}$}

\newcommand{\spitzer}{\textit{Spitzer}}
\newcommand{\chandra}{\textit{Chandra}}

\usepackage{color,graphics,graphicx}

\shorttitle{SOFIA \oiiif\ and \oif\ observations of Cas A} 
\shortauthors{}

\newcommand{\mic}{$\mu$m} 
\slugcomment{}
\shorttitle{SOFIA far-infrared observations of Cas A}

\begin{document}

\title{SOFIA Far-infrared \oiiif\ and \oif\ Observations of 
Dense CO-knots in the Supernova Remnant Cassiopeia A: Multi-phase Ejecta}

\author{ 
Jeonghee Rho\altaffilmark{1},  
Sofia H. J. {Wallstr{\"o}m}\altaffilmark{2, 3},
S{\'e}bastien Muller\altaffilmark{4},
Isabelle Cherchneff\altaffilmark{5},
Dario Fadda\altaffilmark{6},
Olivier Bern{\'e}\altaffilmark{7, 8},
John H. Black\altaffilmark{4},
Alexander G. G. M. Tielens\altaffilmark{9}
} 
\altaffiltext{1}{SETI Institute, 189 N. Bernardo Ave, 
Mountain View, CA 94043; jrho@seti.org}
\altaffiltext{2}{Institute of Astronomy, KU Leuven, Celestijnenlaan 200D bus 2401, 3001 Leuven, Belgium}
\altaffiltext{3}{Institute of Astronomy and Astrophysics, Academia Sinica, 11F of 
Astronomy-Mathematics Building, No.1, Sec. 4, Roosevelt Rd., Taipei 10617, Taiwan}
\altaffiltext{4}{Department of Space, Earth and Environment, Chalmers University of Technology, Onsala Space Observatory, SE-43992 Onsala, Sweden}
\altaffiltext{5}{Instituto de Fisica Fundamental Consejo Superior de Investigaciones Científicas (CSIC),
Serrano 113 bis  28006, Madrid Spain; i.chercheneff@iff.csic.es}
\altaffiltext{6}{SOFIA Science Center/USRA, NASA Ames Research Center, M.S. N232-12, Moffett Field, CA 94035}
\altaffiltext{7}{Université de Toulouse, UPS-OMP, IRAP, 31028 Toulouse, France}
\altaffiltext{8}{CNRS, IRAP, 9 Av. Colonel Roche, BP 44346, 31028 Toulouse Cedex 4, France}
\altaffiltext{9}{Leiden Observatory, Leiden University, PO Box 9513, 2300 RA Leiden, The Netherlands}

\begin{abstract} 

Dense, fast-moving ejecta knots in supernova remnants are prime sites for
molecule and dust formation. We present SOFIA far-IR spectrometer FIFI-LS
observations of CO-rich knots in Cas A which cover a $\sim$1 square arc
minute area of the northern shell, in the [O\,III] 52 and 88\mic\ and
[O\,I] 63\mic\ lines. The FIFI-LS spectra reveal that the line profiles of
[O\,III] and [O\,I] are similar to those of the {\it Herschel} PACS [O\,III] and
CO lines. We find that the [O\,III] maps show very different morphology
than the [O\,I] map. The [O\,III] maps reveal diffuse, large-scale structures
and the ratio of the two [O\,III] lines imply the presence of gas with a
range of density 500 - 10,000 cm$^{-3}$ within the mapped region. In
contrast, the [O\,I] map shows bright emission associated with the dense
CO-rich knots. The 63\mic\ \oif\ line traces cooled, dense post-shocked gas
of ejecta. We find that IR-dominated \oiiif\ emission is from post-shocked
gas based on its morphology, high column density, and velocity profile. We
describe multi-phase ejecta knots, a lifetime of clumps, and survival of dust
in the young supernova remnants.

 
\end{abstract}

\section{Introduction}

Massive stars ($>$8 M$_{\odot}$) eject much of the elements synthesized
over its lifetime back into the interstellar medium (ISM) at the end of its
lifetime. This is one of the main ways in which a galaxy's metallicity is
increased, thereby enriching the next generation of stars and planets
\citep{pagel97}. Supernova are also a main source of kinetic energy for the
ISM, stirring up the gas, providing turbulent support to the ISM and
processing interstellar dust through powerful shocks \citep{tielens05book}.
Instabilities create dense clumps in the ejecta \citep{hammer10} and
hundreds of such dense Fast Moving Knots (FMKs) are commonly observed in
type II supernova remnants (SNRs), containing a substantial fraction of the
supernova ejecta and its kinetic energy \citep{fesen01}. The high density
of these knots is conducive to molecule and dust formation
\citep{sarangi15, sluder18}. However, whether dust can survive the reverse
shock is still not clear \citep{nozawa07, nath08, silvia12, biscaro16}.

Understanding the formation and survival of supernova dust has gotten
additional impetus with the recent {\it Herschel} detection of copious amounts of
dust in SNRs, Cas A (0.1-0.6M$_\odot$), SN 1987A (0.5M$_\odot$) and
G54.1+0.3 (0.1-0.9M$_\odot$) \citep{barlow10, deLooze17, matsuura11,
rho18}, as well the detection of copious amounts of dust in the early
universe \citep{laporte17, spilker18}, which is often ascribed to injection
by supernovae \citep{pei91, cherchneff10}. Understanding the
characteristics of dense knots in SNRs and their interaction with the
reverse shock and eventual dispersion and merging with the ISM is therefore
a key question for the evolution of galaxies and the cosmic history of
dust.

Cassiopeia A (Cas A) is one of the youngest Galactic SNRs with an age of
$\sim$350 yr and its progenitor is believed to be a Wolf-Rayet star with a
mass of 15-25 M$_\odot$ \citep{krause08}. \oif\ observation was reported
by \citet[][DS10]{docenko10} with ISO/LWS. The {\it Herschel Space Observatory}
detected 88\mic\ \oiiif\ lines, and high-J rotational CO lines
\citep[][W13]{wallstrom13}. We report \oif\ and \oiiif\ maps toward the
northern part of Cas A over a $\sim$1 square arc minute with SOFIA
Field-Imaging Far-Infrared Line Spectrometer (FIFI-LS), and this letter is
one of the first observations with FIFI-LS.

\section{Observations}

\begin{figure*}[!ht]
\includegraphics[angle=0,width=18.truecm]{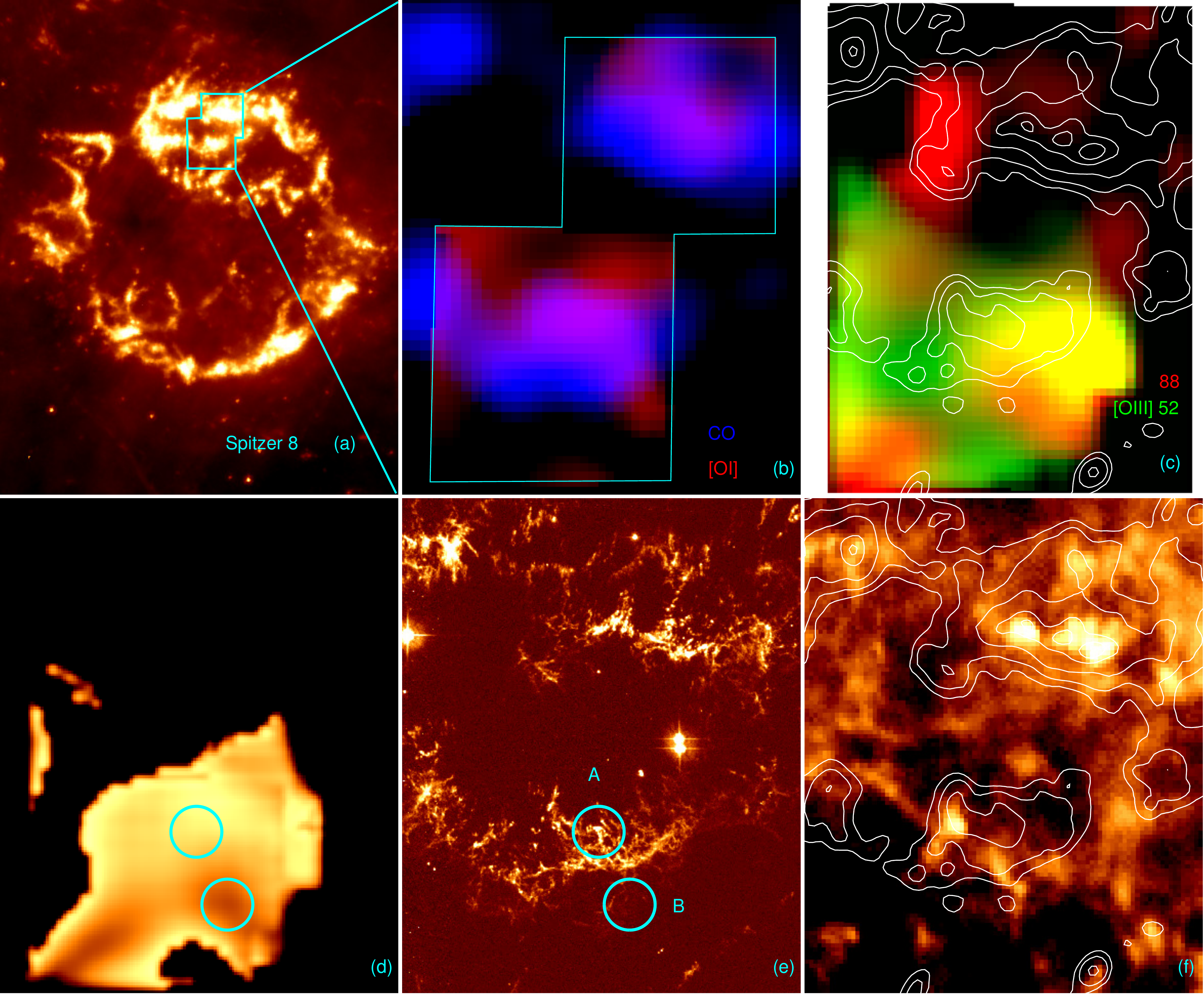}
\caption{(a) The field of view (FOV) of SOFIA FIFI-LS observations (polygon in cyan,
covering 1$'$ area) is marked on \spitzer\ 8\mic\ map of Ar ejecta in Cas A
\citep{rho08}. The image covers 4.7$'$$\times$6.2$'$ centered on R.A.\
$23^{\rm h} 23^{\rm m} 25.82^{\rm s}$ and Dec.\ $+58^\circ$48$^{\prime}
50.42^{\prime \prime}$ (J2000). (b) Zoomed image of Cas A (1$'$$\times$1$'$
FOV, the polygon is the same as in panel (a)) centered on the  SOFIA
FIFI-LS 63\mic\ \oif\ image (the flux ranges 0.035-0.70 Jy/ 1$''$ pixel) is
superposed on {\it Spitzer} CO map at 4.5$\mu$m which was smoothed  to the
same resolution.  The panels (c, d, e) cover the same FOV. (c) Mosaicked
\oiiif\ emission image at 52$\mu$m (in green, the flux ranges 0.1-0.63
Jy/pixel) and 88\mic\ (in red; the flux ranges 0.01-0.08 Jy/pixel). The
white contours are \spitzer\ 8$\mu$m Ar image from \cite{rho08}. (d) The ratio
map of \oiiif\ emission (using the pixels with S/N$>$5) is derived between
52$\mu$m/88$\mu$m emission. The scale of the ratio ranges from $\sim$2 to 9
(brighter color with a higher value) which correspond to a density from 500
to a few times of 10$^4$ cm$^{-3}$ (see the text and
Fig.\,\ref{oiiimodel}). Two circles indicate the \oiiif\
optical-bright/IR-dim ejecta region ($``$Region A" with the ratio
52/88$\mu$m $\sim$8)) and IR-bright/optical-dim ($``$Region B" with the
ratio 52/88$\mu$m $\sim$2) regions marked on the ratio map (d) and {\it HST}
optical \oiiif\ image (e). (f) {\it Chandra} X-ray image is superposed on
\spitzer\ 8$\mu$m contours, showing that X-ray gas is anti-correlated to
the IR-8$\mu$m ejecta in the south although the X-ray ejecta have IR counterparts
in the north.  
}
\label{fifilsmaps}
\end{figure*}

\begin{figure}
\includegraphics[scale=0.6,angle=0,width=8.5truecm]{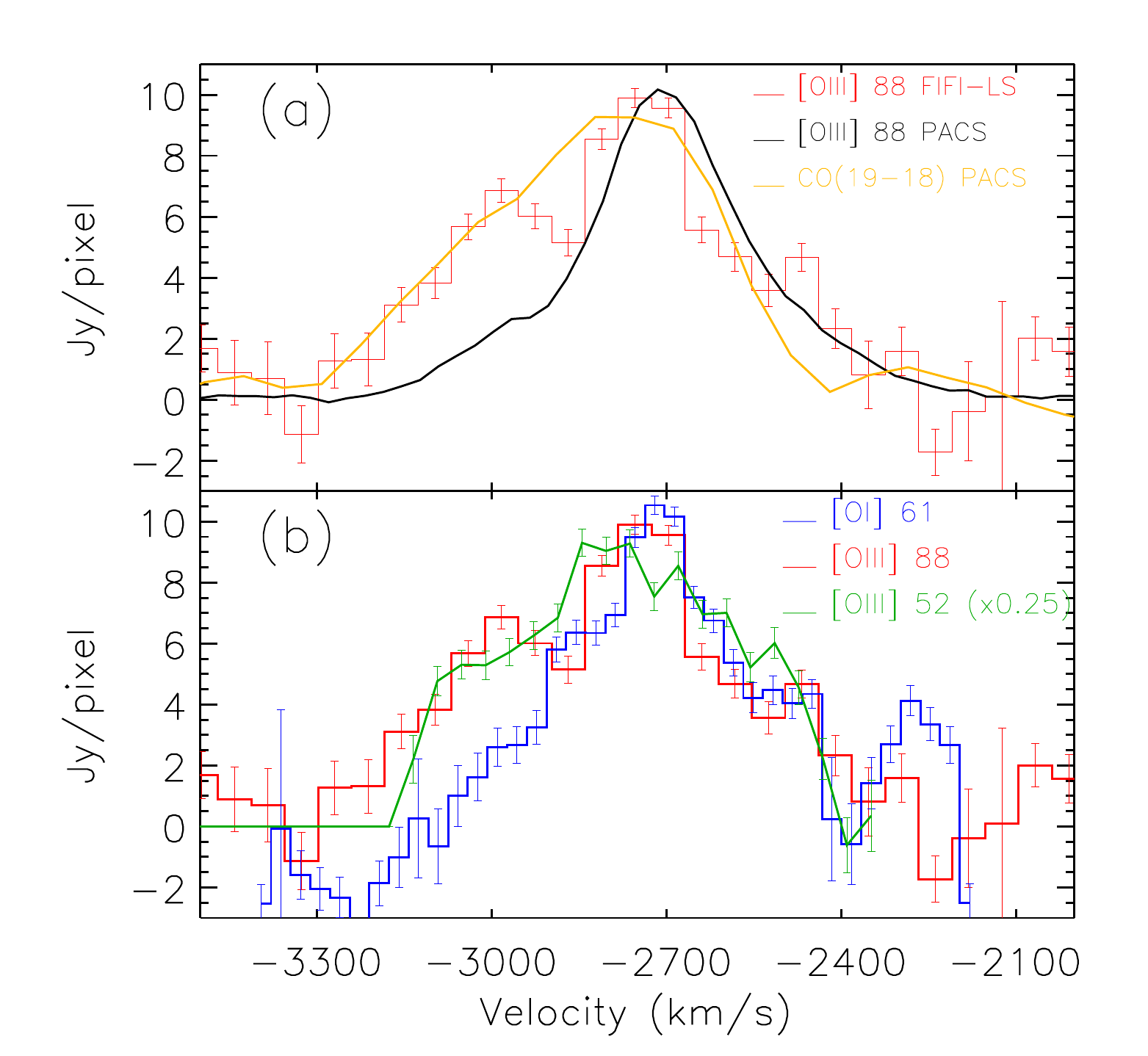}
\caption{The FIFI-LS spectrum of [O~III] at 88\mic\ extracted from an area of 9$''$
pixel (a) is compared with that of {\it Herschel} PACS 88\mic\ \oiiif\ and
CO(19-18) (W13); the spectra show similar line profiles. (b) SOFIA FIFI-LS
spectrum of 63\mic\ \oif\ is compared with the 88 and 52$\mu$m \oiiif\
spectra.}
\label{casafifilsspec}
\end{figure}

We have targeted a region in the north of Cas A with FIFI-LS on the  SOFIA
2.5-m telescope, for the lines of \oiiif\ at 52 and 88$\mu$m and \oif\ at
63\mic. The observations covered two partially overlapping 30$''$x30$''$
regions as shown in Figure~\ref{fifilsmaps}a where two observations of NW
(knot2) and SE (knot1) fields were combined. The SE location (knot1) is
first recognized as Ne-rich crescent-shaped (a.k.a.\,Ne moon) region
\citep{ennis06}, and CO-rich knots \citep[see Fig. 4 of][]{rho12}. The
\oiiif\ at 52$\mu$m and 88$\mu$m, and \oif\ at 63\mic\ observations took
place on 2014 April 26, and additional observations for the \oiiif\ at 52
and 88$\mu$m took place on 2015 October 14 and 21 (PROG\_ID of 02\_0058 and
03\_0051; PI. Tielens). The on-source exposures are 69, 86, and 43 min for
\oif\ 63\mic, \oiiif\ 52 and 88\mic, respectively. A chopping/nodding mode
was used with a chop throw of 300$''$, and FIFI-LS covers a field of view
(FOV) of 30$''$$\times$30$''$ in the lines we present. FIFI-LS spectral
cube has a sampling with a pixel size of 1$''$, and the spatial resolution
of the blue (51-120$\mu$m) channel is 6$''$ \citep{colditz18} because of a better spatial sampling
and more efficient dithering than the default spectroscopy mode of Herschel
PACS (W13). The pointing stability of SOFIA is better than 0.4$''$
\citep{temi18}. The data used is the FIFI-LS pipeline version of 1\_3\_3
produced by SOFIA Science Center.

FIFI-LS calibration uncertainty on the blue channel is at 15\% level, and
line observations are corrected for atmospheric absorption and telluric
features by the SOFIA FIFI-LS team using ATRAN models. We checked the
atmospheric transmission ($>$0.8) curve for each line, and the primary line
structures and fluxes are not significantly affected but it may add some
uncertainty ($<$20\%). The spectral cube was reduced using the Python GUI
software {\it SOSPEX}{\footnote{https://github.com/darioflute/sospex/}}
\citep{fadda18}. We obtained consistent results by using the other data
reduction methods such as
$``$fluxer"{\footnote{http://www.ciserlohe.de/fluxer/fluxer.html}} and our
own IDL extraction routines.

\section{Results}

We extracted FIFI-LS spectra from the central area (a 9$''$x9$''$ area
centered on R.A.\ $23^{\rm h} 23^{\rm m} 24.9^{\rm s}$ and Dec.\
$+58^\circ$50$^{\prime} 03.3^{\prime \prime}$, J2000) of the PACS
observations (W13) as shown in Figure~\ref{casafifilsspec}. The surface
brightness of \oif\ and \oiiif\ lines listed in
Table~\ref{tab:Tfifilslines}A are comparable to those from ISO 
(a beam size of 80$''$, \# 2 of DS10)
and PACS (W13) within a factor of
1.2-2.6.

For detailed comparison, we extracted PACS 88$\mu$m \oiiif\ spectra from
archival data to match the exact extraction area of two different missions
and instruments. We note that recently processed PACS data (version
v14.2.0) yield fainter brightness (in Table 1) than the one by
\cite{wallstrom13}. The FIFI-LS 88$\mu$m \oiiif\ line brightness is 60\%
smaller than that of the PACS spectrum and the line profiles show
400-600 \kms. We use 60\% as our systematic uncertainty of the brightness,
which would not change our main science results considering that the SNR is
a diffuse source.

\begin{table*}
\caption[]{Spectral Line Brightnesses of Observed Lines and Physical Properties of Line Emitting Regions} \label{tab:Tfifilslines}
\begin{center}
\begin{tabular}{llllllll}
\hline \hline
{\bf Table 1A}& Wavelength  &    Wavelength           &  Shift    & FWHM    & FWHM    &  Surface Brightness\\
  Line   & bandwidth($\mu$m)   &center ($\mu$m)        &   (\kms) & ($\mu$m)   &  (\kms) &  (erg\,s$^{-1}$\,cm$^{-2}$\,sr$^{-1}$)   \\ \hline
\oiiif\ & 51.27-51.56 (51.29-51.72)$^a$ & 51.3357$\pm$0.0089 & -2772$\pm$52 & 0.0924$\pm$0.0327 & 540$\pm$190& 1.841$\pm$0.092 $\times$10$^{-3}$   \\
\oiiif\ &87.04-88.05 (87.36-88.34) &87.5445$\pm$0.0150& -2755$\pm$51 &  0.1501$\pm$0.0260& 514$\pm$89 &2.700$\pm$0.117 $\times$10$^{-4}$ \\
\oiiif\ & PACS$^a$: 85.27-89.14 & 87.5594$\pm$0.0100&-2678$\pm$35 &  0.0960$\pm$0.0016& 329$\pm$07 & 4.360$\pm$0.065 $\times$10$^{-4}$\\
\oif\ & 62.41-62.96 (62.43-62.98) & 62.6138$\pm$0.0127 & -2706$\pm$61& 0.0964$\pm$0.0447 &461$\pm$214 & 4.027$\pm$0.328 $\times$10$^{-4}$ \\ 
\hline
\hline
{\bf Table 1B}  &  &&&\\
{Region$^b$} & 5007\AA$^c$ & 52\mic\ &   $R(52/88)$ & $R(5007/52)$ & $R(52/5007)$ & $R(88/5007)$\\
            & (erg\,s$^{-1}$\,cm$^{-2}$\,sr$^{-1}$) &(erg\,s$^{-1}$\,cm$^{-2}$\,sr$^{-1}$) &\\
A (Ejecta)  & 3.65$\pm$0.14 $\times$10$^{-2}$& 17.7$\pm$6.0 $\times$10$^{-4}$   & $\sim$8 & 20$\pm$7 &  0.05 & 0.006 \\ 
B (IR-\oiiif) & 4.60$\pm$0.50 $\times$10$^{-3}$ & 8.2$\pm$4.1 $\times$10$^{-4}$   & $\sim$2 & 5.6$\pm$3.0 & 0.36 & 0.18 \\
\hline
\hline
{\bf Table 1C} & density   & temperature  & column density  & $\Delta$R$^d$  & Einstein A & Pressure  \\ 
Emission &  n(cm$^{-3}$) &  T (K) & (cm$^{-2}$) &  (cm) & (s$^{-1}$) & (cm$^{-3}$\,K) \\ 
IR \oiiif $^e$ & 500 &  6600  &10$^{16}$ & 2$\times$10$^{13}$/$f$ & A(52)=9.8E-5  & $\sim$3$\times$10$^6$ \\
Optical \oiiif $^f$ & 10$^{4}$ & 6000  & 10$^{15}$& 10$^{11}$/$f$ & A(5007)=2E-2 & $\sim$1$\times$10$^8$\\
IR \oif\ & $\sim$10$^{6}$ & ...&2$\times$10$^{15}$ & 2$\times$10$^{9}$/$f$ & A(63)=8.9E-5 &...& \\
CO& 10$^{6}$ & 400, 2000 &5$\times$10$^{17}$ & 5$\times$10$^{11}$/$f$ &... & (5-10)$\times$10$^8$& \\
\hline
\hline
\end{tabular}
\end{center}
{\footnotesize $^a$ The wavelength bandwidth in parenthesis is the coverage for the northwestern region.}\\
{\footnotesize $^b$ Regions A and B are marked in Figure~\ref{fifilsmaps}d and Figure~\ref{fifilsmaps}e.}\\
{\footnotesize $^c$ The optical flux is extinction corrected.}\\
{\footnotesize $^d$ $\Delta$R is the thickness of the emitting region and $f$ is a filling factor.}\\
{\footnotesize $^{e, f}$ ``IR \oiiif" is IR-bright and optical dim \oiiif\ region estimated from $``$Region A",
and $``$Optical \oiiif" is optical-bright and IR dim \oiiif\ region estimated from $``$Region B".}
\end{table*}

We present velocity integrated maps (the velocity range is from -3300 to
-2400 \kms; see Fig.~\ref{casafifilsspec}) of \oif\ 63, \oiiif\ 52 and
88\mic\ maps in Figure~\ref{fifilsmaps}. The 88$\mu$m FIFI-LS map is
consistent with that of PACS although the existing PACS data have only 25
spatial pixels as the region was undersampled. Both 88 and 52\mic\ \oiiif\
maps show very different structure from the [O I] map. The \oif\ 63\mic\
map reveals clumpy structures and bright emission associated with the dense
ejecta and CO-rich knots \cite[][]{rho08} in Figure~\ref{fifilsmaps}b. In
contrast, the \oiiif\ 52 and 88\mic\ maps show mostly diffuse, large-scale
structures and bright emission clearly present outside of the region of
dense ejecta knots in Figure~\ref{fifilsmaps}c. The 63\mic\ \oif\ line is 50\% brighter than the
88$\mu$m \oiiif\ line and 22\% of 52 $\mu$m \oiiif\ brightness, as listed
in Table 1A.

We estimated the ratio of \oiiif\ 52 and 88\mic\ maps since the ratio is a
density indicator (see Fig.\,\ref{oiiimodel} and discussion below).  We
used the pixels with positive values for both 52 and 88 \mic\ emission,
where the NW regions (knot2) within the observed FOV show weak \oiiif\
emission. We find ratios between $\sim$2 and $\sim$9
(Fig.~\ref{fifilsmaps}d).

\section{Discussion}

High-J CO emission  with {\it Herschel} (W13) together with {\it Spitzer},
AKARI, and ground-based IR studies \citep{rho09, rho12} of dense clumps in
Cas A have revealed large column densities (4$\times$10$^{17}$ cm$^{-2}$)
of warm, dense CO gas (W13). The theoretical model of \cite{biscaro14}
suggested that CO molecules are destroyed in the reverse shock but rapidly
reformed in the postshock gas. As a result of the large density contrast
between the dense knots and their surroundings, the reverse shock driven
into the knots is relatively weak and dust can survive in the CO-rich gas.
Morphologically, the 63\mic\ \oif\ emission (Fig.~\ref{fifilsmaps}b) is
closely related to dense ejecta/CO-knot emission. The \oif\ line traces
cooled, dense post-shocked gas of ejecta and the luminosity of \oif\ for
9$''$$\times$9$''$ area is 0.5 L$_\odot$. Because of its high critical
density, the \oif\ emission likely arises from very dense ($\sim$10$^{5-6}$
cm$^{-3}$) post-shock gas, with a high column density of 2$\times$10$^{15}$
cm$^{-2}$ (Table 1A); these physical properties are consistent with the theoretical
shock models \citep[][DS10]{borkowski90}.

\begin{figure}
\includegraphics[scale=0.9,angle=90,width=9.truecm]{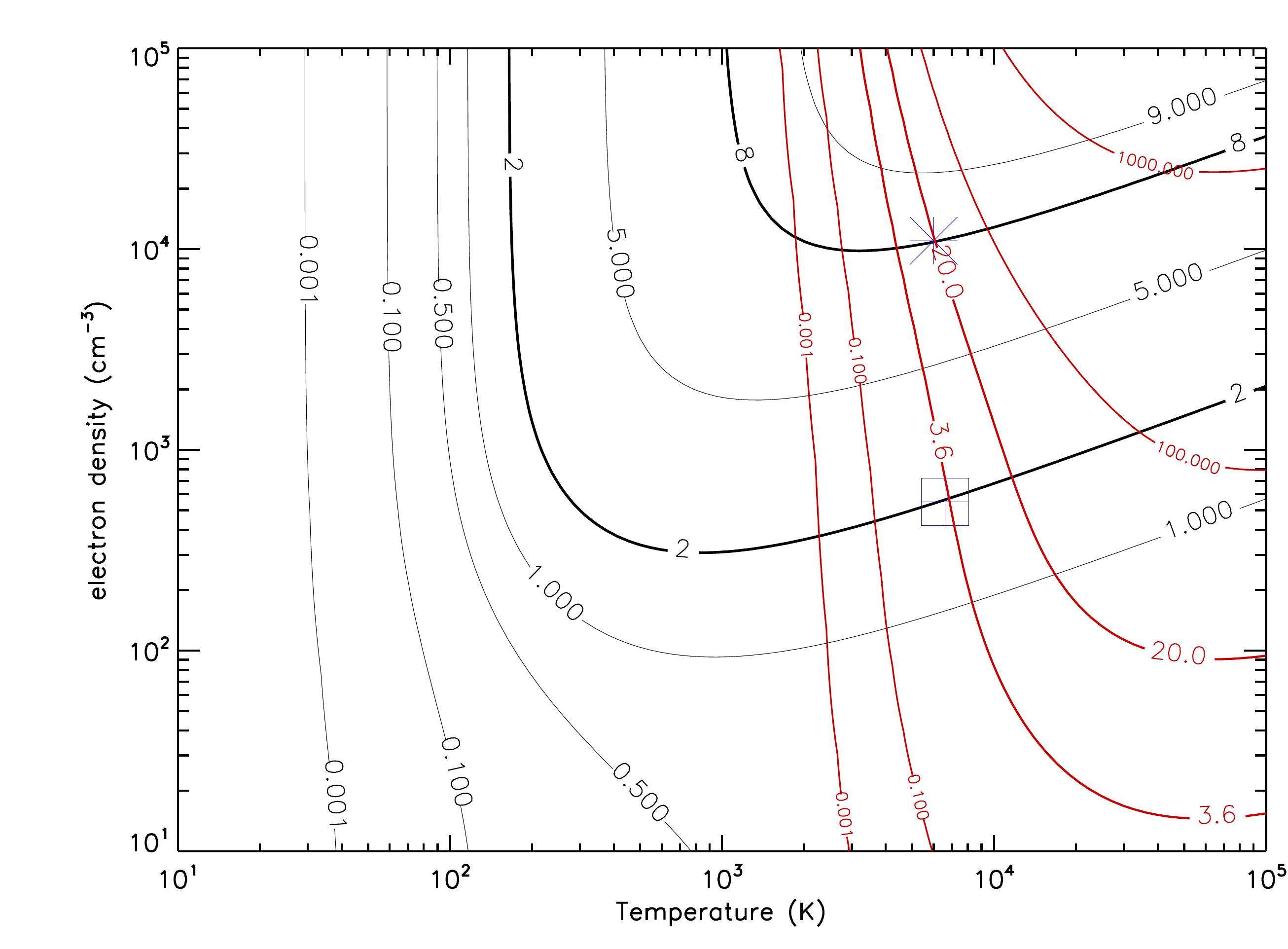}
\caption{Line diagnostic contours of the ratios of IR \oiiif\
52$\mu$m/88$\mu$m, a density indicator (black) and optical to IR ratios of
\oiiif\ between 5007\AA\ and 52\mic\ (red). The observed ratios are marked
with the thick solid lines. The model infers the bright ejecta region has a
density of 10$^{4}$ cm$^{-3}$, and a temperature of 6000\,K (marked as an
asterisk), and the IR \oiiif\ bright region has a density of 550 cm$^{-3}$
and a temperature of 6600\,K (a square).}
\label{oiiimodel}
\end{figure}

Based on the energy levels of the ground-state atomic fine structure, the
ratios of [O III] lines constrain the density and temperature of the
emitting region, independent of the geometry. We generated the set of
\oiiif\ IR line ratio between 52\mic\ and 88$\mu$m with a grid of density
and temperature \citep[see][]{reach00} and the line diagnostic
model \citep[see also][]{osterbrock88} is shown in Figure \ref{oiiimodel}. 
The IR \oiiif\ line ratio between \oiiif\ 52$\mu$m and 88$\mu$m ranges from
2 to 9, and the IR \oiiif\ line diagnostic implies a density range of 300 -
5$\times$10$^4$ cm$^{-3}$ (and a temperature $>$ 100\,K). The ratio map in
Figure~\ref{fifilsmaps}d corresponds to a density map, indicating a higher
density at the ejecta and neighboring region, but the ratio map differs
from the ejecta map.

The \oiiif\ far-IR morphology of 52\mic\ and 88\mic\ emission in
Figure~\ref{fifilsmaps}c differs from the dense ejecta emission seen in
optical \oiiif, infrared Ar ejecta, and 63\mic\ \oif\ emission. The optical
\oiiif\ emission at 5007\AA\ shows bright knots at the ejecta position, and
the optical \oiiif\ is faint outside the ejecta (Figure \ref{fifilsmaps}e).
The {\it HST} optical \oiiif\ emission is found near the reverse shock and
shocked gas from a cooling post-shock region \citep{fesen01, morse04}.

What is the origin of far-IR \oiiif$?$ \cite{itoh81} estimated 70\% of
52\mic\ \oiiif\ is from shocked gas using a collisional ionization
equilibrium (CIE) model, and the model by \cite{sutherland95} predicts
bright \oiiif\ below 20,000\,K. To understand the origin and physical
condition of IR-emitting \oiiif\ emission, we selected two regions, one
(Region A, see Figs.~\ref{fifilsmaps}d, \ref{fifilsmaps}e) with the high
ratio of 52/88$\mu$m ($R(52/88)$) of $\sim$8 which coincides with dense
ejecta (centered on R.A.\ $23^{\rm h} 23^{\rm m} 25.11^{\rm s}$ and Dec.\
$+58^\circ$50$^{\prime} 02.8^{\prime \prime}$), and the other (Region B)
with the low $R(52/88)$ of $\sim$2 which is outside the ejecta emission
(centered on R.A.\ $23^{\rm h} 23^{\rm m} 24.64^{\rm s}$ and Dec.\
$+58^\circ$49$^{\prime} 54.22^{\prime \prime}$). $``$Region A" is the
compact dense ejecta region, optical-[O III] bright and CO/[O I] emitting
region, and $``$Region B" is IR-[O III] bright region. We have measured the
brightness of \oiiif\ 5007\AA, 52$\mu$m and 88\mic\ for a 6$''$ circular
aperture. In Table 1B, the brightness and the ratios are listed. We used
{\it HST} ACS F475W image which contains both 5007 and 4959\AA\ lines
\citep{fesen06} and we assumed 3/4 of optical emission is from 5007\AA\
\oiiif\ line and the extinction is A$_v$ = 5.7 mag \citep{hurford96}. When
we compare the line ratios of $R(52/88)$ and $R(52/5007)$ with the line
diagnostics in Figure~\ref{oiiimodel}, IR-[O III] bright regions (Region B)
are dominated by gas with an electron density and temperature of (n$_e$, T)
= (500~cm$^{−3}$, 6000\,K). Optical-[O III] bright, dense knots (Region A)
are dominated by gas with (10$^4$~cm$^{−3}$, 6600\,K). The densities are
much less than the densities from the CO observations (10$^6$ cm$^{-3}$,
W13) or the density required to collisionally excite the [O\,I] line.

We compare the optical and IR \oiiif\ with \chandra\ X-ray map. The optical
\oiiif\ emission is well correlated with the IR dense (Ar) ejecta from
\spitzer\ IRAC 8\mic\ map. The X-ray gas appears outside the dense optical
ejecta (Region A) as shown in Figures~\ref{fifilsmaps}e and
~\ref{fifilsmaps}f and is in some degree associated with IR-\oiiif\ bright
region. Indications of ejecta knot mass stripping in Cas A have been
reported \citep{fesen11}.

\begin{figure}
\includegraphics[scale=1.0,angle=0,width=8.2truecm]{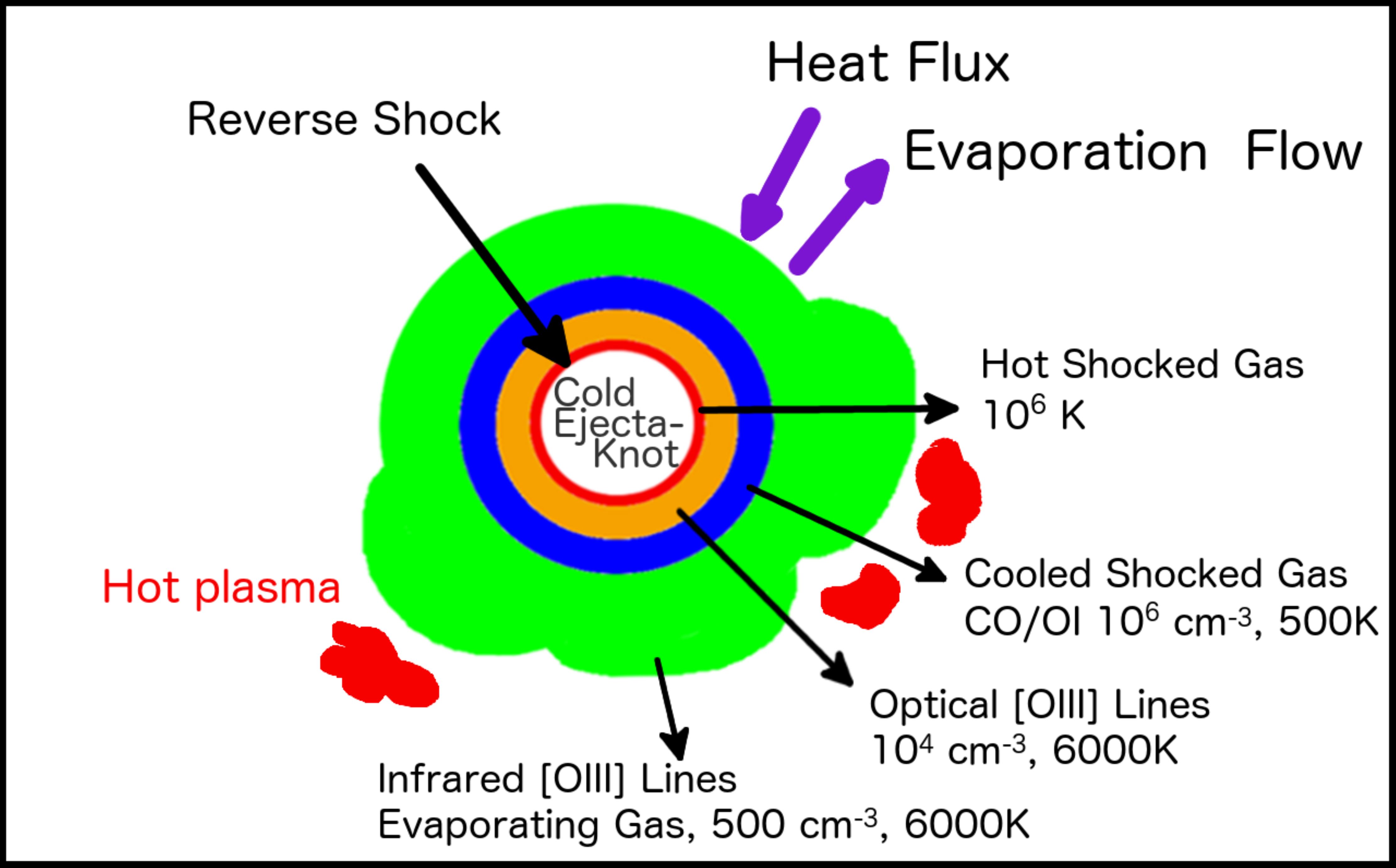}
\caption{Cartoon of CO-rich ejecta knot with multi-phase \oif/CO,  IR \oiiif,
optical \oiiif\ and X-ray emitting layers. The marked physical conditions
are only dominant emission for each. The patches in red are denser gas in
X-rays. 
}
\label{cartoon}
\end{figure}

We estimated the oxygen column density and abundance of oxygen. The column
density is ${4\pi}~I /({A~h\nu})$ for each line, where $I$ is the surface
brightness, $A$ is the line Einstein $A$ coefficient, $h$ is the Planck
constant, and $\nu$ is the frequency. We summarize the physical properties
of IR \oiiif, optical \oiiif, IR \oif, and CO emitting regions in Table 1C.
The column densities of 52 and 88\mic\ \oiiif\ lines are of an order of
10$^{16}$ cm$^{-2}$ and are comparable to those of 63\mic\ \oif\ line and
optical emission.

Shock models predict bright [O\,III] emission arising from the preshock gas
photo-ionized by the shock emission and the postshock cooling and
recombining gas. Such regions are called Photo-Ionized Regions (PIR). It is
tempting to associate the low density gas bright in the IR but not the
optical [O\,III] lines (Region B) with the preshock PIR while the denser
optically bright [O\,III] emission (Region A) would naturally arise from
the postshock gas \citep{fesen11, docenko08}. Given the uncertainties in the shock models,
the observed IR-[O\,III] intensities and derived column
densities, densities and temperatures are in good agreement with calculated
post-shock models \citep[][DS10]{sutherland95}. However, the morphology of
the [O\,III] emission is not in agreement with this interpretation. The PIR
gas should be in the region interior to the reverse shock; e.g., deep
within the FMK. Observationally, this gas is very diffuse and surrounds the
dense clumps and, hence, it should be postshock gas. So, rather than PIR
gas, we surmise that the IR bright/optically dim [O\,III] emitting gas is
associated with the evaporating gas in the surfaces of the shocked clumps
driven by electron conduction \citep{borkowski90}. Another evidence that IR
\oiiif\ emission is post-shocked gas comes from the velocity information.
The velocity profiles of \oiiif\ are the same as that from shocked gas of
63\mic\ \oif\ line. Unshocked gas using \spitzer\ \oivf\ line (from the
central region) indicates the velocity for the gas toward us peaks at -2000
\kms. We expect to observe a difference in velocity between shocked and
unshocked gas and this is not the case for IR \oiiif\ lines as shown in
Figure \ref{casafifilsspec}.

The observed CO and [O\,I] emission is most naturally associated with
postshock gas that has had time to cool, recombine and form molecules
\citep{biscaro14}.  This gas is directly related therefore to the [O\,III]
emitting gas. The pressure of the optically bright [O\,III] emitting gas is
$\sim$10$^8$ cm$^{-3}$\,K as summarized in Table 1C. This is a little less
than the pressure derived from the CO observations of 5$\times$10$^8$
cm$^{-3}$\,K (W13).  The agreement in morphology supports this
interpretation. In contrast, the low density, IR bright [O\,III] emitting
gas has a gas pressure of $\sim$3$\times$10$^6$ cm$^{-3}$\,K; much less
than the pressure of the postshock gas. Momentum conservation in the
evaporating gas requires $\rho\, v^2\, =\, P_{ps}$ with $P_{ps}$ the
pressure in the the postshock gas and where we have neglected the gas
pressure in the flow. This results in a flow velocity of ~3 km/s, or --- as
expected --- about the sound speed, of an oxygen gas at 6000\,K.  

The mass loss rate of the clumps as measured by the evaporation flow is then, 
\begin{equation}
\dot{M}\, =\, 4\pi R^2\, nm_O v\, \simeq \, 3\times 10^{-5}
\qquad \text{M$_{\odot}$/yr},
\end{equation}
where $R$ is the radius of the emitting region ($2\times 10^{17}$ cm), 
$n$ the density (500 cm$^{-3}$) and $m_O$ the oxygen mass.
The lifetime of the clumps is
\begin{equation}
\tau_c\, =\, \frac{4\pi\, R^2Nm_O}{4\pi\, R^2nm_O v}\, = \, \frac{N_{\text{CO}}}{X_{\text{CO}}n v}\, =\, 300\, X_{\text{CO}^{-1}} ~\text{yr}, 
\end{equation}
with $N$ the column density of the clump. This column density is at least
equal to the postshock column density which is equal to
$N_{\text{CO}}/X_{\text{CO}}$ with N and X the CO column density and
abundance. Observationally, CO is more abundant than O in the postshock gas
and this would imply a lifetime of the clumps at least comparable to the
lifetime of the remnant. For calculated CO abundances  of 10$^{-2}$
\citep[][]{biscaro14}, the expected lifetime is some 3$\times$10$^4$ yr and
the knots may well survive the SNR phase. Finally, we note that the optical
knots show variations on a timescale of tens of years \citep{fesen11} and
we attribute this to variable geometric structure associated with the
heavily corrugated surface of the clumps, reflecting the effects of the
Rayleigh Taylor instabilities.

Figure 4 illustrates the structure of the region in a cartoon, identifying
the different zones present in this region. The knots are oxygen-rich gas
traversed by a $\sim$200 \kms\ (reverse) shock. The hot ($\sim$10$^6$\,K)
gas layer immediately behind the shock is very thin. The optical bright
[O\,III] gas represents the cooling and recombining postshock gas, once the
gas has cooled down and recombined, CO is formed and it as well as [O\,I]
are the dominant coolant of this gas. The evaporating surface layers are
bright in the IR [O\,III] lines but not the optical lines. Eventually, this
evaporating gas will mix into the dominant X-ray emitting gas that fills
the interior of the remnant \citep{fesen11, patnaude14}. This flux of
evaporating gas is driven by heat conduction due to the fast electrons in
the X-ray plasma.

The CO/\oif\ emitting layers are postshock gas (e.g.,\,material processed
by a $\sim$200 \kms\ shock) and now cooled down. For the CO/\oif\ emitting
layers that have a density of 10$^{5-6}$ cm$^{-3}$, the velocity in the
knot (V$_k$) ranges 6 - 20 \kms\ where C-shock is dominated. Grain
sputtering requires a velocity greater than 50 \kms\ \citep{draine95,
tielens94}. IR-\oiiif\ emitting regions have a temperature of $\sim$6000\,K
where no thermal sputtering occurs \citep[$<$10$^5$\,K,][]{biscaro16} and
have a counterpart of cold dust when we examined {\it Herschel} 70 and
160\mic\ maps, indicating that dust destruction is not significant. Recent
estimates by \cite{nath08} and \cite{micelotta16} show only 1-20\% of
grains are sputtered in SN ejecta and the cooling in SN ejecta is rapid due
to metal enriched ejecta. This long sputtering timescale as well as the
constant reloading of the hot gas with dust from evaporating FMKs may be
the reasons that we still observe a significant amount of dust in the hot
gas processed by the reverse shock in Cas A. In the end, only those clumps
that $``$survive" the whole SNR expansion phase or that travel beyond the
strong shock plowing into the ISM will contribute to seeding the ISM with
supernova dust.

In conclusion, SOFIA FIFI-LS observations reveal the morphology of the
[O\,I] and [O\,III] line emission in the SNR Cas~A. 
The \oif\ traces cooled, dense shocked gas and is related to CO
emission and dense ejecta. The 52 and 88\mic\ \oiiif\ emission 
reveals a component bright in the optical [O\,III] lines with a density of
10$^4$ cm$^{-3}$ and temperature of 6000\,K. This emission is coincident with
the dense knots and we associate this with the cooling and recombining
postshock gas layer before the gas has cooled down to 500\,K and CO has
formed. The IR [O\,III] maps reveal the region with $n$$\sim$500 cm$^{-3}$
and $T$$\sim$6000\,K. This gas envelopes the dense knots and we attribute
this emission to gas evaporating from the dense knots due to the heat
conduction by fast electrons.

\acknowledgements
We are grateful to Rob Fesen for providing {\it HST} images, and Kazik Borkowski
for providing helpful comments on the earlier draft. Results in this paper
are based on observations made with the NASA/DLR Stratospheric Observatory
for Infrared Astronomy (SOFIA). SOFIA is jointly operated by the
Universities Space Research Association, Inc. (USRA), under NASA contract
NAS2-97001, and the Deutsches SOFIA Institut (DSI) under DLR contract 
50-OK-0901 to the University of Stuttgart. Financial support for this work in
part was provided by NASA through award SOF-03-0051, issued by USRA.

\bibliography{msrefs}

\end{document}